\documentstyle[epsf]{mn2e}

\newcommand{\bG}{\mbox{\boldmath$G$}}
\newcommand{\bT}{\mbox{\boldmath$T$}}
\newcommand{\bl}{\mbox{\boldmath$l$}}

\title[Warped discs around black holes]
{The evolution of a warped disc around a Kerr black hole}

\author[S. H. Lubow et al.]
{S. H. Lubow,$^{1,2}$ G. I. Ogilvie,$^{1,2}$ and J. E. Pringle$^{1,2}$\\
$^1$Space Telescope Science Institute, 3700 San Martin Drive,
Baltimore, MD 21218, USA\\
$^2$Institute of Astronomy, University of Cambridge, Madingley Road,
Cambridge CB3 0HA}

\date{\today}

\begin{document}

\maketitle

\begin{abstract}
  We consider the evolution of a warped disc around a Kerr black hole,
  under conditions such that the warp propagates in a wavelike manner.
  This occurs when the dimensionless effective viscosity, $\alpha$,
  that damps the warp is less than the characteristic angular
  semi-thickness, $H/R$, of the disc.  We adopt linearized equations
  that are valid for warps of sufficiently small amplitude in a
  Newtonian disc, but also account for the apsidal and nodal
  precession that occur in the Kerr metric.  Through analytical and
  time-dependent studies, we confirm the results of Demianski \&
  Ivanov, and of Ivanov \& Illarionov, that such a disc takes on a
  characteristic warped shape.  The inner part of the disc is not
  necessarily aligned with the equator of the hole, even in the
  presence of dissipation.  We draw attention to the fact that this
  might have important implications for the directionality of jets
  emanating from discs around rotating black holes.
\end{abstract}

\begin{keywords}
  accretion, accretion discs -- black hole physics -- hydrodynamics --
  waves
\end{keywords}

\section{Introduction}

Accretion discs are found around black holes in the centres of active
galaxies and also in galactic X-ray binary stars.  These systems may
form in such a way that the angular momentum vectors of the disc and
of the black hole are not parallel.  As a particle in a tilted orbit
around a rotating black hole undergoes Lense--Thirring precession at a
rate dependent on the radius of the orbit, so a tilted disc
experiences a gravitomagnetic torque that tends to twist and warp the
disc.  If the disc is able to communicate a warping disturbance in a
diffusive manner as a result of its pressure and viscosity, it may
adopt a characteristic warped shape in which the plane of the disc
undergoes a smooth transition from one plane to another in the
vicinity of a certain radius (Bardeen \& Petterson 1975; Kumar \&
Pringle 1985; Scheuer \& Feiler 1996).  At large radius the disc is
essentially flat and its plane is determined by the total angular
momentum vector of the disc.  At small radius the disc is essentially
flat and lies in the equatorial plane of the black hole.  Over a
longer time-scale the disc and hole tend towards mutual alignment
(Scheuer \& Feiler 1996; Natarajan \& Pringle 1998).

However, there are circumstances in which a warping disturbance
propagates in a wavelike, rather than diffusive, manner.  This occurs
in a Keplerian disc when the dimensionless viscosity parameter
$\alpha$ (Shakura \& Sunyaev 1973) is smaller than the angular
semi-thickness $H/R$ of the disc (Papaloizou \& Lin 1995; Wijers \&
Pringle 1999; see the discussion by Pringle 1999); in a non-Keplerian
disc the propagation is also wavelike, although it is dispersive
(Ogilvie 1999).  The evolution of a disc around a rotating black hole
under these conditions has received less attention.  On the basis of
analytical calculations, Demianski \& Ivanov (1997) and Ivanov \&
Illarionov (1997) found that the disc can adopt a steady warped shape
in which the tilt angle is an oscillatory function of radius.  More
recently, Nelson \& Papaloizou (2000) conducted smoothed particle
hydrodynamic simulations of accretion discs around rotating black
holes under a variety of assumptions.  They did not find the tilt
oscillations, but found that the inner part of the disc was aligned
with the equatorial plane of the hole.

In this paper we re-examine the shape of a warped disc around a Kerr
black hole, under conditions such that the warp propagates in a
wavelike manner.  We explore the reasons for the existence of a steady
wavelike solution, and discuss the expected wavelength, amplitude and
phase of the oscillations in disc tilt.  Using time-dependent
calculations of the linearized equations, we study the process by
which such a steady solution is established.  Finally, we discuss the
potentially important implications of the shape of the disc for the
directionality of jets emanating from discs around black holes.

\section{Wavelike propagation of warps}

The propagation of warping disturbances of small amplitude in a nearly
Keplerian and nearly inviscid disc has been discussed by Papaloizou \&
Lin (1995), Masset \& Tagger (1996), Demianski \& Ivanov (1997),
Pringle (1999) and Lubow \& Ogilvie (2000).  These authors differ in
respect of the notations and assumptions they adopt, but all agree
that bending waves propagate approximately non-dispersively under
these conditions, with a wave speed that is somewhat less than the
average sound speed of the disc.

We consider disturbances of a steady disc of surface density
$\Sigma(R)$ and vertically integrated pressure $P(R)$ defined by
\begin{equation}
  \Sigma=\int\rho\,{\rm d}z,\qquad
  P=\int p\,{\rm d}z,
\end{equation}
where $(R,\phi,z)$ are cylindrical polar coordinates.  Let
$\Omega(R)$, $\kappa(R)$ and $\Omega_z(R)$ be the orbital angular
velocity, the epicyclic frequency and the vertical oscillation
frequency associated with circular orbits at radius $R$ from the
central object.  After an integration by parts, and using the vertical
hydrostatic equilibrium of the disc, we find
\begin{equation}
  P=\int z\left(-{{\partial p}\over{\partial z}}\right){\rm d}z=
  \Omega_z^2\int\rho z^2\,{\rm d}z=\Omega_z^2\Sigma H^2,
\end{equation}
where $H(R)$ is an effective density scale-height.  In the case of a
vertically isothermal disc, $H$ is equal to the usual Gaussian
scale-height.

A simple way to describe a warped disc is as a continuous set of
tilted circular rings, and to define a unit tilt vector $\bl(R,t)$
normal to the plane of the ring of radius $R$ at time $t$.  When the
disc is only slightly tilted out of the $xy$-plane, $l_z\approx1$.  As
shown by Lubow \& Ogilvie (2000), the equations governing warping
disturbances of small amplitude in a nearly Keplerian and nearly
inviscid disc may be written in the form
\begin{equation}
  \Sigma R^2\Omega{{\partial\bl}\over{\partial t}}=
  {{1}\over{R}}{{\partial\bG}\over{\partial R}}+\bT,
  \label{dldt}
\end{equation}
\begin{equation}
  {{\partial\bG}\over{\partial t}}-
  \left({{\Omega^2-\kappa^2}\over{2\Omega}}\right)\bl\times\bG+
  \alpha\Omega\bG=
  {{PR^3\Omega}\over{4}}{{\partial\bl}\over{\partial R}}.
  \label{dgvecdt}
\end{equation}
Equation (\ref{dldt}) is derived from the vertical component of the
equation of motion, and can be understood as expressing the
conservation of horizontal angular momentum: $2\pi\bG(R,t)$ is the
horizontal internal torque in the disc, and $\bT(R,t)$ is the
horizontal external torque, per unit area, acting on the disc.
Equation (\ref{dgvecdt}) is derived from the horizontal components of
the equation of motion, and determines the internal torque $\bG$.  The
internal torque is mediated by shearing epicyclic motions, driven by
the radial pressure gradients that are set up when a stratified disc
is warped.  These motions undergo apsidal precession when
$\kappa\ne\Omega$, and also viscous decay, here described by a
dimensionless viscosity parameter $\alpha$.

The parameter $\alpha$ appears in equation (\ref{dgvecdt}) only to
quantify the rate, $\alpha\Omega$, at which the shearing epicyclic
motions associated with the warp decay.  Although this term is derived
from a viscous model, numerical simulations of the interaction of
shearing epicyclic motions with three-dimensional magnetorotational
turbulence (Torkelsson et al. 2000), and an analytical closure model
of such turbulence (Ogilvie 2002), both suggest that small-amplitude
shearing epicyclic motions decay exponentially in time.  The measured
rate is in fact through to be somewhat smaller than $\alpha\Omega$ if
$\alpha$ is calibrated in the usual way with reference to the
$R\phi$-component of the turbulent stress.

It is often more convenient to adopt a complex representation in which
$W=l_x+{\rm i}l_y$ and $G=G_x+{\rm i}G_y$.  Then
\begin{equation}
  \Sigma R^2\Omega\left[{{\partial W}\over{\partial t}}-
  {\rm i}\left({{\Omega^2-\Omega_z^2}\over{2\Omega}}\right)W\right]=
  {{1}\over{R}}{{\partial G}\over{\partial R}},
  \label{dwdt}
\end{equation}
\begin{equation}
  {{\partial G}\over{\partial t}}-
  {\rm i}\left({{\Omega^2-\kappa^2}\over{2\Omega}}\right)G+
  \alpha\Omega G=
  {{PR^3\Omega}\over{4}}{{\partial W}\over{\partial R}}.
  \label{dgdt}
\end{equation}
Note that the external torque is related to the non-sphericity of the
potential, and therefore to the difference between $\Omega^2$ and
$\Omega_z^2$. In addition, $W(R,t)$ represents the tilt of the disc at
each radius in the sense that $W(R,t) = \beta(R,t) \exp [i \gamma
(R,t)]$, where $\beta$ is the amplitude of the local tilt and $\gamma$
is the azimuth (cf. Pringle 1996).

These equations are known to be valid when the quantities $W$ and $G$
are sufficiently small, and vary on a length-scale long compared to
$H$ and on a time-scale long compared to $\Omega^{-1}$.  In deriving
them it is also assumed that the quantities $|1-\Omega_z^2/\Omega^2|$,
$|1-\kappa^2/\Omega^2|$ and $\alpha$ are of order $H/R$ or smaller.

In the absence of viscosity, the dispersion relation associated with
this system, for wavelike solutions with a rapidly varying phase
factor $\exp\int({\rm i}\omega\,{\rm d}t-{\rm i}k\,{\rm d}R)$, is
\begin{equation}
  \left[\omega-\left({{\Omega^2-\Omega_z^2}\over{2\Omega}}\right)\right]
  \left[\omega-\left({{\Omega^2-\kappa^2}\over{2\Omega}}\right)\right]=
  {{1}\over{4}}k^2H^2\Omega_z^2.
  \label{dispersion}
\end{equation}
This shows that, in an exactly Keplerian disc, the warp propagates as
a non-dispersive wave with wave speed $H\Omega_z/2$, which is equal to
half the isothermal sound speed in the case of a vertically isothermal
disc.  The presence of nodal and/or apsidal precession introduces
dispersion, and forbids wave propagation for frequencies between
$(\Omega^2-\Omega_z^2)/2\Omega$ and $(\Omega^2-\kappa^2)/2\Omega$
(Fig.~1).

\begin{figure}
  \centerline{\epsfbox{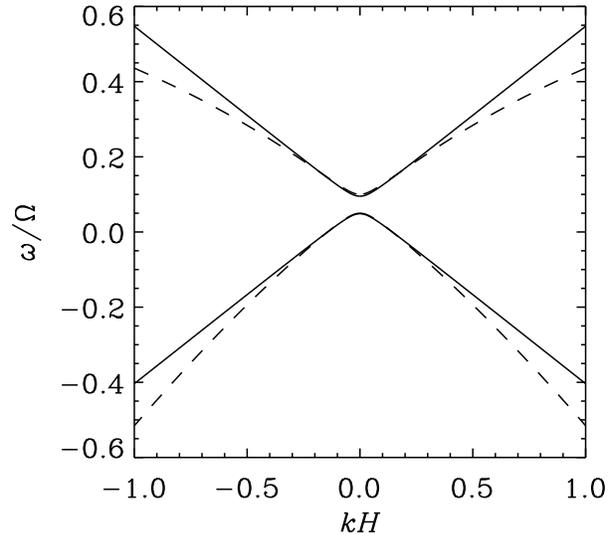}}
  \caption{Dispersion relation for bending waves in the case of
    prograde nodal and apsidal precession with $\Omega_z/\Omega=0.95$
    and $\kappa/\Omega=0.9$, according to equation (\ref{dispersion})
    (solid lines) and equation (\ref{lp}) (dashed lines).}
\end{figure}

It is interesting to compare this dispersion relation with that
derived by Lubow \& Pringle (1993), which describes the full set of
wave modes of a vertically isothermal disc.  Ogilvie \& Lubow (1999)
slightly extended that work to allow for the possibility that
$\Omega_z\ne\Omega$, and found that, for a disc undergoing isothermal
perturbations ($\gamma=1$),
\begin{equation}
  \left({{\hat\omega^2-\Omega_z^2}\over{\Omega_z^2}}\right)-
  \left({{\hat\omega^2}\over{\hat\omega^2-\kappa^2}}\right)k^2H^2=n,
\end{equation}
where $\hat\omega=\omega-m\Omega$, $m$ is the azimuthal mode number
and $n$ is the vertical mode number.  For the case $m=1$, $n=0$, which
corresponds to a tilt or warp, we find
\begin{eqnarray}
  \lefteqn{\left[\omega-
  \left({{\Omega^2-\Omega_z^2+\omega^2}\over{2\Omega}}\right)\right]
  \left[\omega-\left({{\Omega^2-\kappa^2+\omega^2}\over{2\Omega}}
  \right)\right]}&\nonumber\\
  &&={{1}\over{4}}k^2H^2\Omega_z^2\left(1-{{\omega}\over{\Omega}}\right)^2.
  \label{lp}
\end{eqnarray}
Unlike equation (\ref{dispersion}), this dispersion relation is valid
when the quantities $|1-\Omega_z^2/\Omega^2|$, $|1-\kappa^2/\Omega^2|$
and $kH$ are of order unity.  The two equations clearly agree well in
the limit of low frequency, $|\omega/\Omega|\ll1$, and agree exactly
when $\omega=0$ (see Fig.~1).

The case $\omega=0$ is of some interest, as it relates to warped discs
that are independent of time.  When both nodal and apsidal precession
are present, and are either both prograde or both retrograde, the
dispersion relation for $\omega=0$ indicates that the spatial
structure of the warp is oscillatory in character, its radial
wavenumber being given by
\begin{equation}
  k^2H^2={{(\Omega^2-\Omega_z^2)(\Omega^2-\kappa^2)}\over
  {\Omega^2\Omega_z^2}}.
  \label{k2h2}
\end{equation}
An oscillatory solution of this kind may be interpreted as a bending
wave having zero phase velocity but non-zero group velocity.

Conversely, when the nodal and apsidal precession are in opposite
senses, a steady warp is spatially evanescent.  This situation is
qualitatively similar to the case when the viscosity is large and the
warp satisfies a diffusion equation.  It applies when the orbital,
epicyclic and vertical frequencies are derived from an axisymmetric
Newtonian gravitational potential that satisfies Laplace's equation in
the mid-plane of the disc,
\begin{equation}
  \kappa^2-2\Omega^2+\Omega_z^2=0,
\end{equation}
since this implies
\begin{equation}
  \hbox{sgn}\,(\Omega^2-\Omega_z^2)=-\hbox{sgn}\,(\Omega^2-\kappa^2).
\end{equation}
However, this need not hold in the metric of a black hole.

\section{The steady shape of a warped disc around a black hole}

The orbital, epicyclic and vertical frequencies around a Kerr black
hole satisfy (Kato 1990)
\begin{equation}
  \Omega^{-1}=\left({{GM}\over{c^3}}\right)\left(r^{3/2}+a\right),
  \label{kerr1}
\end{equation}
\begin{equation}
  {{\Omega^2-\kappa^2}\over{\Omega^2}}=6r^{-1}-8ar^{-3/2}+3a^2r^{-2},
  \label{kerr2}
\end{equation}
\begin{equation}
  {{\Omega^2-\Omega_z^2}\over{\Omega^2}}=4ar^{-3/2}-3a^2r^{-2},
  \label{kerr3}
\end{equation}
where $-1<a<1$ is the dimensionless angular momentum parameter of the
hole ($a<0$ implying a retrograde disc) and $r$ is the radius in
units of $GM/c^2$.  For a prograde disc ($a>0)$ we have the ordering
$\kappa<\Omega_z<\Omega$ at any radius, corresponding to a situation
as illustrated in Fig.~1.

Fig.~2 shows the radial wavelength, $\lambda=2\pi/k$, associated with
the steady wavelike shape of a prograde warped disc around a Kerr
black hole.  To a first approximation, i.e. considering only the first
terms in equations (\ref{kerr1})--(\ref{kerr3}), we have
\begin{equation}
  {{\lambda}\over{H}}\approx{{\pi r^{5/4}}\over{(6a)^{1/2}}}.
\end{equation}
The wavelength is shorter for more rapidly rotating black holes, and
decreases as the marginally stable circular orbit is approached, but
is always at least a few times $H$.  This suggests that these wavelike
solutions are physically realizable and may be described with some
confidence by theories of long-wavelength bending waves.

\begin{figure}
  \centerline{\epsfbox{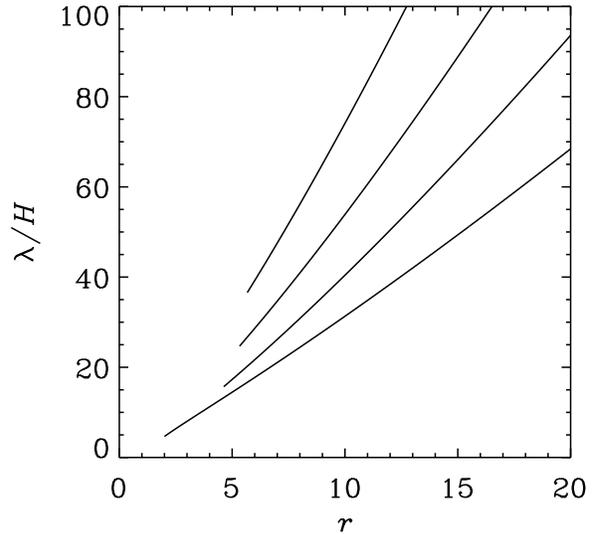}}
  \caption{Ratio of the radial wavelength of the steady warp to the
    effective density scale-height, plotted against the distance from
    the black hole, expressed in units of $GM/c^2$.  The four curves
    relate to Kerr black holes with $a=0.1$ (top), $a=0.2$, $a=0.4$
    and $a=1$ (bottom).  The inner radius plotted corresponds to the
    marginally stable circular orbit in each case.}
\end{figure}

A more accurate description of the steady wavelike shape can be
obtained by setting the time-derivatives to zero in equations
(\ref{dwdt}) and (\ref{dgdt}).  This results in a second-order
ordinary differential equation for the tilt variable $W(R)$,
\begin{equation}
  {{{\rm d}}\over{{\rm d}R}}\left[\left({{PR^3\Omega^2}\over
  {\Omega^2-\kappa^2+2{\rm i}\alpha\Omega^2}}\right)
  {{{\rm d}W}\over{{\rm d}R}}\right]+\Sigma R^3(\Omega^2-\Omega_z^2)W=0.
  \label{steady}
\end{equation}
If we take the first-order approximations from equations
(\ref{kerr1})--(\ref{kerr2}), we have
\begin{equation}
  {{{\rm d}}\over{{\rm d}R}}\left[\left({{\Sigma H^2}\over
  {3r^{-1}+{\rm i}\alpha}}\right)
  {{{\rm d}W}\over{{\rm d}R}}\right]+8ar^{-3/2}\Sigma W=0.
  \label{approx}
\end{equation}
Under the further assumptions that $\Sigma\propto R^{-1/2}$ and
$H\propto R$, this equation is equivalent to that studied by Ivanov \&
Illarionov (1997), who first identified the steady wavelike solutions.

A WKB treatment of equation (\ref{steady}) in the absence of viscosity
also indicates that the radial variation of the amplitude of the
oscillations in $W$ is proportional to
\begin{equation}
  \left[\left({{\Omega^2-\Omega_z^2}\over{\Omega^2-\kappa^2}}\right)
  \Sigma^2H^2R^6\Omega^4\right]^{-1/4}
\end{equation}
and therefore, to a first approximation, proportional to
\begin{equation}
  {{R^{1/8}}\over{(\Sigma H)^{1/2}}}.
\end{equation}
The amplitude may either increase or decrease as one approaches the
inner radius, depending on the properties of the disc, but is likely
to do so slowly.

More specifically, if $\Sigma\propto R^{-2\sigma}$ and $H/R=\epsilon
r^{h-1}$ with $\epsilon={\rm constant}$ and
$h>\sigma>-{\textstyle{{1}\over{4}}}$, equation (\ref{approx}) in the
absence of viscosity has a solution in Bessel functions,
\begin{equation}
  W=x^\nu\left[C_1J_\nu(x)+C_2Y_\nu(x)\right],
\end{equation}
where
\begin{equation}
  \nu={{h-\sigma}\over{h+{\textstyle{{1}\over{4}}}}},\qquad
  x=\left({{24a}\over{\epsilon^2}}\right)^{1/2}
  {{r^{-(h+1/4)}}\over{h+{\textstyle{{1}\over{4}}}}}.
\end{equation}
At large $R$ ($x\ll1$) the second Bessel function is dominant and the
tilt tends to a constant value,
\begin{equation}
  W\to W_\infty=-{{2^\nu}\over{\pi}}\Gamma(\nu)C_2.
\end{equation}
At small $R$ ($x\gg1$) the Bessel functions are oscillatory and
\begin{eqnarray}
  \lefteqn{W\sim x^\nu\left({{2}\over{\pi x}}\right)^{1/2}
  \left[C_1\cos\left(x-{{\nu\pi}\over{2}}-{{\pi}\over{4}}\right)\right.}
  &\nonumber\\
  &&\left.\qquad
  +C_2\sin\left(x-{{\nu\pi}\over{2}}-{{\pi}\over{4}}\right)\right].
\end{eqnarray}
Note that the amplitude of $W\propto x^{\nu-1/2}\propto R^{1/8}(\Sigma
H)^{-1/2}$ as quoted above.  In the special case considered by Ivanov
\& Illarionov (1997), $\sigma={\textstyle{{1}\over{4}}}$ and $h=1$, so
$\nu={\textstyle{{3}\over{5}}}$ and the amplitude of $W\propto
R^{-1/8}$.  A different special case occurs when
$\nu={\textstyle{{1}\over{2}}}$, since then
\begin{equation}
  W=\left({{2}\over{\pi}}\right)^{1/2}(C_1\sin x-C_2\cos x)
\end{equation}
exactly.

While the constant $C_2$ is fixed by the tilt at large radius, $C_1$
is determined by the inner boundary condition.  If this is such that
the torque vanishes at the inner radius $R=R_{\rm in}$ (usually at the
marginally stable circular orbit), then ${\rm d}W/{\rm d}R=0$ there.
In the special case $\nu={\textstyle{{1}\over{2}}}$, this determines
the solution as
\begin{equation}
  W=W_\infty{{\cos(x_{\rm in}-x)}\over{\cos x_{\rm in}}},
  \label{cos_solution}
\end{equation}
provided that $\cos x_{\rm in}\ne0$.  This solution can be understood
as a standing wave consisting of a superposition of ingoing and
outgoing bending waves (having zero phase velocity but non-zero group
velocity).  The condition of vanishing torque at the inner radius
reflects the ingoing bending waves.  Note that in this case the warp
consists of pure tilt, without twisting (i.e. $\gamma={\rm
  constant}$).

The steady shape of the warp can be interpreted in terms of a resonant
wave-launching process (e.g. Goldreich \& Tremaine 1979). However, in
the present case, the resonance is at $R= \infty$ because the forcing,
caused by the steady disc tilt at large radius, occurs at zero
frequency. In addition, the region of maximum forcing, close to the
black hole, is far removed from the resonance (cf. Lubow \& Ogilvie
2000, where the resonance is at $R=0$ and the forcing is concentrated
in the outer part of the disc). As a consequence, the forcing
effectively occurs at an intermediate radius, typically where the wave
executes its outermost wavelength ($k R \sim x \sim 1$). The wave is
launched as a trailing ($k > 0$) inwardly propagating wave, having
group velocity less than or of order the sound speed (see also the
discussion by Nelson \& Papaloizou 2000).

\section{Time-dependent calculations}

We now demonstrate how these equations can be applied to the
time-dependent behaviour of a disc in the neighbourhood of a rotating
black hole.  The simple application we have in mind here is the
Newtonian approximation to Lense--Thirring precession around a rotating
black hole. In this case, we can write approximately (Demianski \&
Ivanov 1997; Nelson and Papaloizou 2000; Section~3 above)
\begin{equation}
\eta \equiv {{\kappa^2-\Omega^2}\over{2\Omega^2}} =
- \frac{3}{2} \frac{R_{\rm s}}{R},
\end{equation}
and
\begin{equation}
\zeta \equiv {{\Omega_z^2-\Omega^2}\over{2\Omega^2}} =
- \frac{a}{\sqrt{2}} \left(\frac{R_{\rm s}}{R}\right)^{3/2},
\end{equation}
where $R_{\rm s} = 2GM/c^2$ is the Schwarzschild radius.

In the numerical calculations we choose to solve for the quantities
$A(R,t)$ and $D(R,t)$, which define the radial velocity perturbation
and the enthalpy perturbation according to
\begin{equation}
  u_R'=Az\,{\rm e}^{{\rm i}\phi},\qquad
  w'=Dz\,{\rm e}^{{\rm i}\phi},
\end{equation}
where the real part is understood.  These are simply related to $W$
and $G$ by $W=-D^*/R\Omega_z^2$ and $G={\textstyle{{1}\over{2}}}\Sigma
H^2R^2\Omega A^*$.  To be explicit, we use the equations (\ref{dwdt})
and (\ref{dgdt}) to solve for $A$ and $D$. Thus,
\begin{equation}
\label{ANK0diss}
\frac{\partial A}{\partial t} = - \frac{D}{R}
-\frac{1}{2}\frac{\partial D}{\partial R}
+ i \eta \Omega A
- \alpha \Omega A,
\end{equation}
and

\begin{equation}
\label{DNK1}
\frac{\partial D}{\partial t} =
- \frac{c_{\rm s}^2}{2} \left[ \frac{1}{\Sigma H^2 R^{1/2}}
 \frac{\partial}{\partial
 R}(\Sigma H^2 R^{1/2} A) \right] +i \zeta \Omega D.
\end{equation}

We consider a disc which is inclined to the spin axis of the hole at
large radius.  For ease of comparison with the analytic results we
consider the particular case in which steady shape of the disc is
described by Bessel functions of order $\frac{1}{2}$, and so can be
written in terms of trigonometric functions.  Thus we assume the disc
with inner radius at $R=R_{\rm in}$ to have $\Sigma = \Sigma_{\rm in}
(R/R_{\rm in})^{-1/2}, H = H_{\rm in} (R/R_{\rm in})^{3/4} , \Omega =
\Omega_{\rm in} (R/R_{\rm in})^{-3/2}$, and thus (in the vertically
isothermal case) a sound speed $c_{\rm s} = c_{\rm s,in} (R/R_{\rm
  in})^{-3/4}$. If we let

\begin{equation}
\eta = \eta_{\rm in} (R/R_{\rm in})^{-1},\qquad
\zeta = \zeta_{\rm in} (R/R_{\rm})^{-3/2},
\end{equation}
then the steady solution of the equations, with $\alpha = 0 $, is

\begin{equation}
D = \beta_{\infty} R^{-2} \frac{\cos[\lambda(R_{\rm in}^{-1} - R^{-1})]}
{\cos[\lambda / R_{\rm in}]},
\end{equation}
and
\begin{equation}
A = \frac{{\rm i} \beta_{\infty} \lambda}{2 \eta_{\rm in} \Omega_{\rm in}
R_{\rm in}} (R/R_{\rm in})^{-3/2} \frac{\sin[\lambda(R_{\rm in}^{-1} -
R^{-1})]}{\cos[\lambda / R_{\rm in}]},
\end{equation}
where
\begin{equation}
\lambda^2 \equiv \frac{4 \eta_{\rm in} \zeta_{\rm in} R_{\rm
in}^4}{H_{\rm in}^2},
\end{equation}
and $\beta_{\infty}$ is the tilt angle of the disc as $R \rightarrow
\infty$. This solution is chosen to ensure that there is zero torque
($A=0$) at the inner boundary.

\subsection{Numerical results}

The numerical method is to take $A$ to be defined at the $N$
logarithmically distributed grid points (typically $N=1001$ or
$2002$), and to take $D$ to be defined at the half-grid points. We
take $A$ and $D$ to be defined both at full time points and at the
intermediate half-time points. $A$ is updated from: (a) $D$ at the
intermediate time-point using straightforward numerical differencing
for the radial derivative, (b) $A$ at the intermediate time-point for
the precession term (leapfrog), and (c) $A$ at the previous full
time-point for the dissipative term (Euler method).  Similarly $D$ is
updated from: (a) $A$ at the intermediate time-point using
straightforward numerical differencing for the radial derivative, and
(b) $D$ at the intermediate time-point for the precession term
(leapfrog). Boundary conditions are required only for $A$, and we take
$A = 0$ at the boundaries, which corresponds to a zero torque
condition at the inner and outer disc radii.

For numerical convenience, we take units in which $R_{\rm in} = 1$,
$\beta_{\infty} = 1$, and $\Omega_{\rm in} = 1$. We then choose the
other parameters to produce a Newtonian approximation to a rotating
black hole for which the angular momentum parameter is $a =
\frac{2}{3} (4 - \sqrt{10} ) \approx 0.5585$ and for which the
marginally stable circular orbit is at $R_{\rm in} = 2 R_{\rm s}$.
This implies that $\eta_{\rm in} = - \frac{3}{4}$ and $\zeta_{\rm in}
= - \frac{1}{6} (4 - \sqrt{10}) \approx -0.1396$. We then choose
$H_{\rm in}/R_{\rm in} = [(4 - \sqrt{10})/8 \pi^2]^{1/2} \approx
0.1030$ to ensure that $\lambda = 2 \pi R_{\rm in}$. This gives a
steady solution (with viscosity zero, i.e. $\alpha = 0$) of the form:

\begin{equation}
D =  R^{-2} \cos[2 \pi (1 - R^{-1})].
\end{equation}

This solution has the property that the tilt angle $\beta(R) \equiv
|DR^2|$ is unity at $R = 1$, at $R = 2$, and as $R \rightarrow
\infty$, and is zero at $R=4$ and at $R=4/3$. As a check, this
solution was tested numerically by using it as initial input on a grid
from $R_{\rm in} = 1$ to $R_{\rm out} = 30$, using $N=1001$
logarithmically spaced grid points. No evolution was detected.

We then investigated time-dependent behaviour, by considering the
effect of taking a disc which is initially aligned with the spin axis
of the hole, and tilting the outer parts. Thus as an initial condition
we take $D R^2 = 0$ for $R \leq 18$, $D R^2 = \frac{1}{2} ( 1 + \sin
[\pi (R - 20)/4])$ for $18 \leq R \leq 22$, and $D R^2 = 1$ for $R
\geq 22$. To consider a reasonable physical situation we add a small
amount of dissipation and take $\alpha = 0.05$. We take $A = 0$
initially, which satisfies the inner and outer boundary conditions and
also corresponds to the correct solution for a disc with uniform tilt
around a Keplerian point mass. Thus the initial condition corresponds
to a discontinuity in the disc tilt at radius $R = 20$ smoothed over a
distance of $\Delta R = 4$. We integrated the equations for a time of
$t = 20,000$ using a grid with $N=2002$ extending from $R = 1$ to $R =
900$.

\begin{figure}
  \centerline{\epsfbox{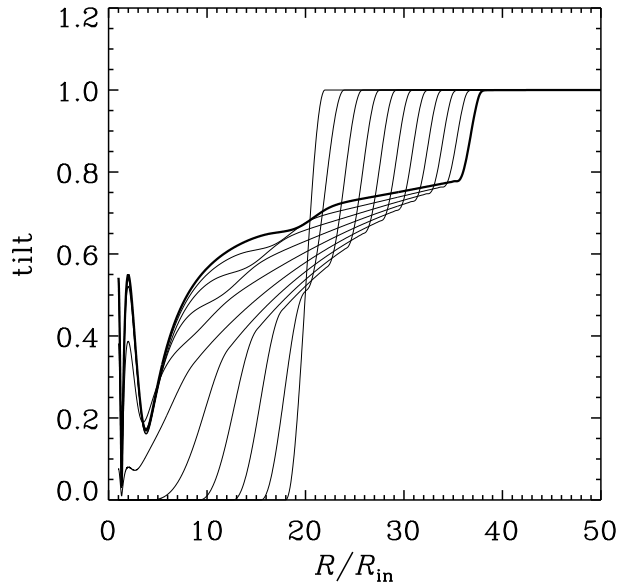}}
  \caption{The tilt of the disc is shown as a function of radius
   at 11 equally spaced times $t = 0, 400, 800, \ldots, 4,000.$ The
  tilt at the final time, $t = 4,000$, shown in bold, shows that the
  steady state solution has already been established at radii $R \la 8$.}
\end{figure}

\begin{figure}
  \centerline{\epsfbox{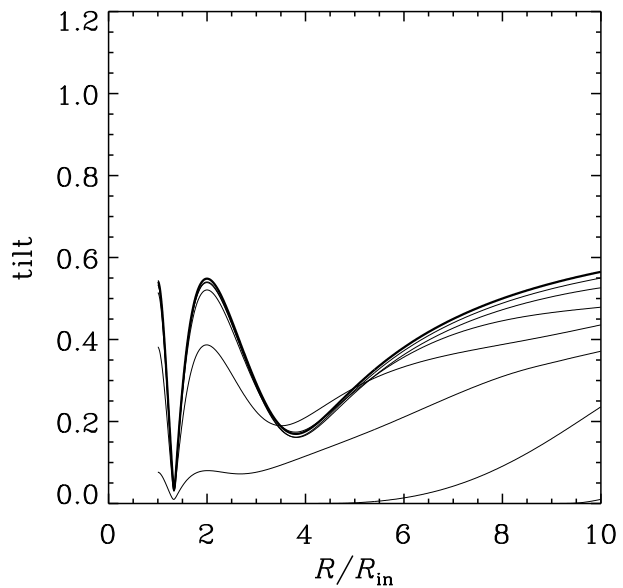}}
  \caption{The tilt of the disc is shown as a function of radius is
    shown for the inner disc regions at 11 equally spaced times $t =
    0, 400, 800, \ldots, 4,000.$ The tilt at the final time, $t =
    4,000$, shown in bold, shows that the steady state solution has
    already been established at radii $R \la 8$.}
\end{figure}

\begin{figure}
  \centerline{\epsfbox{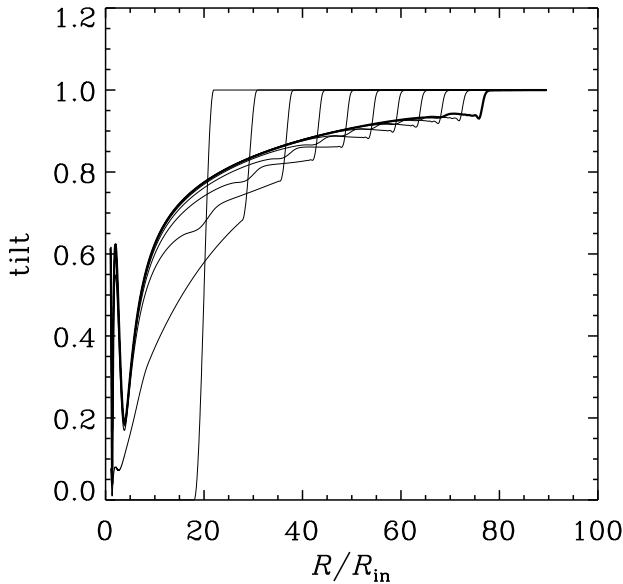}}
  \caption{The tilt of the disc is shown for the inner disc regions at
    11 equally spaced times $t = 0, 2000, 4000, \ldots, 20,000.$ The
    tilt at the final time, $t = 20,000$, shown in bold, shows that
    the steady state solution has already been established at radii $R
    \la 65$.}
\end{figure}

In Fig.~3 we show the initial evolution up to time $t = 4,000$. As
expected the discontinuity in the tilt initially propagates inwards
and outwards in a wavelike fashion. The outward propagation continues
throughout the calculation. The inward propagation of the tilt
interacts with the precessional effects at the hole and by the time $t
= 4,000$ the steady state shape of the disc close to the hole is
essentially established (Fig.~4). It takes the form of a steady warp,
together with a slight twist (caused by the small amount of viscosity)
as predicted by Ivanov \& Illarionov (1997). The interaction of the
inwardly propagating warp with the inner disc regions generates a
reflected wave which propagates outwards. Thus towards the end of the
computation ($t = 20,000$), shown in Fig.~5, at which time the initial
outwardly propagating warp wave is approaching the outer edge of the
grid, followed closely by the reflection of the initially inwardly
propagating warp wave, the steady warped disc solution has been
established over about half the grid.

\section{Discussion}

We have considered the time-dependent evolution of a warped
centrifugally supported disc in the low-viscosity limit ($\alpha \la
H/R$) in which warp propagation is wavelike rather than diffusive. We
note that this is relevant to discs in which the viscosity is low,
such as protoplanetary discs, and/or in which the disc thickness is
large, such as non-radiative or high-luminosity discs around black
holes.  We make use of the linearized perturbation equations, which
require that the tilt angle of the warp be sufficiently small.

As an illustration we apply the resulting equations to a Newtonian
approximation of a moderately thin ($H/R \approx 0.1$), low viscosity
($\alpha = 0.05 \la H/R$) disc around a Kerr black hole with angular
momentum parameter $a \approx 0.5$. The disc in this case is subject
to both apsidal and nodal precession, and the evolution equations we
derive are similar to those derived by Demianski \& Ivanov (1997) in
the limit of low viscosity. Ivanov \& Illarionov (1997) demonstrated
that these equations give rise to a steady state in which the disc
takes up a characteristic warped shape with the precession induced by
the potential being balanced by wavelike stresses in the disc. We
confirm this result, and we use our time-dependent equations to
demonstrate how such a steady state is set up, starting from the
initial condition of a disc whose inner parts are aligned with the
spin axis of the hole, and whose outer parts are tilted at a fixed
inclination. We show that, once the outer tilt has propagated (in a
wavelike fashion) to the central regions of the disc, the steady-state
configuration becomes rapidly established. The adjustment of disc
shape at the inner radii is propagated outwards through the disc at
the usual local wave propagation speed.

Another possible approach to describing the steady shape of a warped
disc around a black hole is based on the work of Ogilvie (1999), who
developed a non-linear theory of bending waves in accretion discs.
The derivation is based on certain ordering assumptions that eliminate
some of the time-derivatives from the problem.  In the time-dependent
case, this theory is applicable to the generic, `non-resonant' case
and not to the special case of discs that are both nearly Keplerian
and nearly inviscid.  However, in the case of steady warps, the theory
of Ogilvie (1999) is applicable to all regimes and has the advantages
of being fully non-linear, and of including the effect of the
accretion flow.  We do not pursue this idea here, but note that for an
inviscid disc with a steady warp, the theory implies that
\begin{equation}
  {\bf0}={{1}\over{R}}{{{\rm d}\bG}\over{{\rm d}R}}+\bT,
\end{equation}
\begin{eqnarray}
  \lefteqn{\bG={{PR^3\Omega^2}\over{2(\Omega^2-\kappa^2)}}
  \,\bl\times{{{\rm d}\bl}\over{{\rm d}R}}}&\nonumber\\
  &&\times\left[1+{{(6+\gamma)}\over{2(3-\gamma)}}
  \left({{\Omega^2}\over{\Omega^2-\kappa^2}}\right)|\psi|^2+
  O(|\psi|^4)\right],
\end{eqnarray}
where $\gamma$ is the adiabatic exponent and $|\psi|=R|{{{\rm
      d}\bl}/{{\rm d}R}}|$ is a dimensionless measure of the amplitude
of the warp.  In the linear regime these equations agree with
equations (\ref{dldt}) and (\ref{dgvecdt}) with the time-derivatives
set to zero.  The theory indicates that non-linear effects become
noticeable when $|R\,{\rm d}W/{\rm d}R|^2$ is comparable to
$(\Omega^2-\kappa^2)/\Omega^2$.  This can be expected to occur in the
inner part of the disc unless the tilt at large radius is sufficiently
small, i.e. $\beta_\infty \ll 1$.  The effect of the non-linearity is
likely to be to increase the wavelength of the warp and restrict its
amplitude.

Other authors have cautioned against taking seriously the possibility
of the disc warp having a substantial amplitude in the centre of the
disc.  Nelson \& Papaloizou (2000) did not find such warps in their
non-linear SPH computations and attribute this failure to a
presumption that ``non-linear effects lead to the damping of short
wavelength [warps], and thus cause the alignment of inner disc regions
in which the tilt amplitude would otherwise change rapidly on small
length-scales''. Although this may be the case, we note that for the
case we consider, with $H/R \sim 0.1$, the radial length-scales of the
warps are generally at least of order the local radius, $R$, (Fig.~2)
and are not small compared to the local disc scale $H$. We note
further that once the radial length-scale of the warp variation
becomes short, the dispersive nature of the warp waves is likely to
become significant and may delay the onset of non-linearity of the
waves.  In addition, in their original paper, Ivanov \& Illarionov
(1997) point out that although the formal computations predict a disc
warp at inner radii, such predictions might be nullified if the warp
itself gives rise to turbulence within the disc, and so a high local
effective viscosity. They suggest that the oscillatory vertical shear
within the disc caused by the warp (Papaloizou \& Pringle 1983) might
become shear unstable. Gammie, Goodman \& Ogilvie (2000) have
investigated this possibility, and show that instability occurs
through a parametric effect and is likely to set in when $|A|\ga 30
\alpha\Omega$.

We have not discussed the interesting question of the net torque
between the disc and the black hole, or the time-scale for mutual
alignment, under conditions such that the warp propagates in a
wavelike manner.  The torque may be considered to be exerted in
launching the steady train of inwardly propagating bending waves,
which carries a certain flux of angular momentum.  Unlike the case of
resonantly launched waves in problems of discs subject to periodic
tidal forcing (Goldreich \& Tremaine 1979), the torque cannot be
simply expressed in terms of the properties of the disc in the
neighbourhood of a certain radius.  Furthermore, if the viscosity is
small enough that the waves reach the inner radius and reflect from
it, the torque that would cause a mutual alignment is partially or
completely cancelled.  For example, in the case
$\nu={\textstyle{{1}\over{2}}}$ leading to the solution
(\ref{cos_solution}) in the absence of viscosity, the integrated
horizontal torque is given in term of an integral over the disc of the
relevant components of the local torque $\bT(R,t)$. Writing $T =
T_x + iT_y$, the total horizontal torque is
\begin{eqnarray}
  2\pi\int_{R_{\rm in}}^\infty T\,R\,{\rm d}R&\propto&
  {{{\rm i}W_\infty}\over{\cos x_{\rm in}}}
  \int_0^{x_{\rm in}}\cos(x_{\rm in}-x)\,{\rm d}x\nonumber\\
  &\propto&{\rm i}W_\infty\tan x_{\rm in}.
\end{eqnarray}
Rather than causing a mutual alignment, this torque causes a slow
mutual precession of the disc and black hole.  The direction and
magnitude of the precession depend sensitively on the parameters of
the disc.

Finally, we note that if the outer disc and the black hole are
misaligned by more than $90\degr$ so that the disc can be considered
retrograde, the steady wavelike solution is replaced by an evanescent
solution.  This occurs because the nodal and apsidal precession are in
opposite senses (see equations~\ref{k2h2}, \ref{kerr2}
and~\ref{kerr3}).  In that case the steady shape of the disc may be
expected to be qualitatively similar to that envisaged by Bardeen \&
Petterson (1975), with the inner disc aligned with the equator of the
black hole but rotating in the opposite sense.  We illustrate this in
Fig.~6.

\begin{figure}
  \centerline{\epsfbox{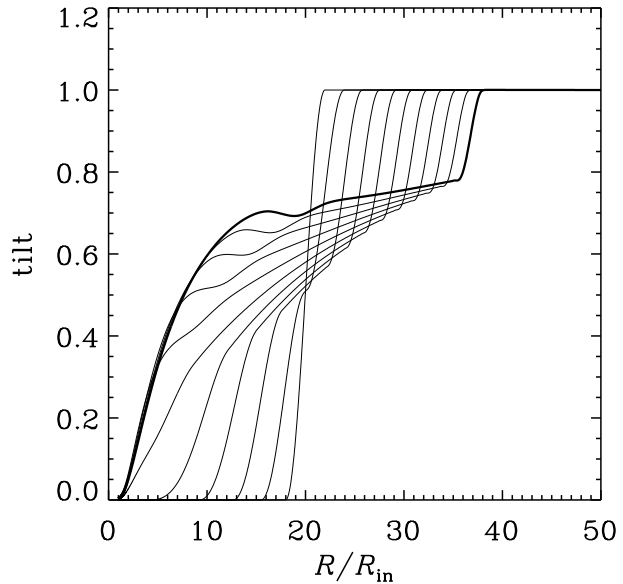}}
  \caption{The result of a calculation identical to that of Fig.~3, but
    in which the sign of the nodal precession $\zeta$ is reversed.
    This corresponds to the physical situation in which the disc and
    hole rotate in opposite senses.}
\end{figure}

\section{Conclusion}

The result that, in low-viscosity discs around a Kerr black hole, the
inner parts of the disc are not necessarily aligned with the black
hole, as found by Ivanov \& Illarionov (1997), is a general one.
Indeed, depending on the disc properties, it is possible for the inner
disc to be tilted at a greater angle to the hole than the outer parts
of the disc. In addition, because the inner disc shape depends
sensitively on the radial dependence of disc properties (such as
surface density and disc thickness), a change (for example) in the
accretion rate can give rise to a change in the inner disc warp, even
without changing the tilt of the outer disc. These results contrast
with the usual finding (e.g. Nelson \& Papaloizou 2000) and/or
assumption (e.g.  Natarajan \& Pringle 1998) that the inner regions of
the disc align with the equator of the hole. In the discussion above,
we have noted above that confirmation of these results awaits a proper
calculation using full general relativity, as well as an assessment of
possible non-linear, dispersive and parametric effects. Nevertheless,
it is evident that the results presented here could have important
implications for the directions in which jets might emanate from
accreting spinning black holes. Indeed, if the region responsible for
direction of jet collimation is at several radii from the hole, which
is likely to be the case for relativistic jets, then the Newtonian
approximations applied above may be adequate to confirm the effect,
even if the very inner regions of the disc are indeed aligned by the
fully relativistic effects close to the hole. A lack of correlation
between the inner and outer disc tilts could provide one explanation
of the finding by Kinney et al. (2000) that the directions of jets
from low luminosity AGN appear to be uncorrelated with the disc plane
of the host spiral galaxies.

\section*{Acknowledgments}

JEP is grateful to STScI for hospitality and for continued support
under its Visitor Program.  GIO acknowledges the support of the Royal
Society through a University Research Fellowship, and of NASA through
grant no. NAG5-10732.

\end{document}